\documentclass[submitting]{nst}

\usepackage{url}
\usepackage{xurl} % 更好的URL断行
\usepackage{subfigure,dcolumn}
\usepackage[T2A,T1]{fontenc}
\usepackage[russian,english]{babel}
\usepackage{lineno}

\usepackage{listings}
\usepackage{multirow}

\begin{document}
\nolinenumbers

\title{Topmetal-L: A Low Noise Charge-Sensitive Pixel Sensor for POLAR-2/LPD}
\par
\thanks{This work is supported by the National Natural Science Foundation of China (Grant Nos. 12027803, U1731239, 12133003), the National Key R\&D Program of China (Grant Nos. 2024YFA1611700, 2023YFE0117200), the Innovation Project of Guangxi Graduate Education (No. YCBZ2025045), and the Guangxi Talent Program (“Highland of Innovation Talents”).}

\author{Li-Rong Xie}
\affiliation{School of Physical Science and Technology, Guangxi University, Nanning 530004, China}

\author{Shi-Qiang Zhou}
\affiliation{PLAC, Key Laboratory of Quark and Lepton Physics (MOE), Central China Normal University, Wuhan, Hubei 430079, China}

\author{Di-Fan Yi}
\affiliation{School of Physical Science and Technology, Guangxi University, Nanning 530004, China}
\affiliation{School of Physical Science, University of Chinese Academy of Sciences, Beijing, 100049, China}

\author{Huan-Bo Feng}
\affiliation{School of Physical Science and Technology, Guangxi University, Nanning 530004, China}

\author{Zu-ke Feng}
\affiliation{School of Physical Science and Technology, Guangxi University, Nanning 530004, China}

\author{Dong Wang}
\affiliation{PLAC, Key Laboratory of Quark and Lepton Physics (MOE), Central China Normal University, Wuhan, Hubei 430079, China}

\author{Chao-Song Gao}
\affiliation{PLAC, Key Laboratory of Quark and Lepton Physics (MOE), Central China Normal University, Wuhan, Hubei 430079, China}

\author{Kai Chen}
\affiliation{PLAC, Key Laboratory of Quark and Lepton Physics (MOE), Central China Normal University, Wuhan, Hubei 430079, China}

\author{Xiang-Ming Sun}
\affiliation{PLAC, Key Laboratory of Quark and Lepton Physics (MOE), Central China Normal University, Wuhan, Hubei 430079, China}

\author{En-Wei Liang}
\affiliation{School of Physical Science and Technology, Guangxi University, Nanning 530004, China}

\author{Hong-Bang Liu}
\email[Corresponding author, ]{Hong-Bang Liu, liuhb@gxu.edu.cn}
\affiliation{School of Physical Science and Technology, Guangxi University, Nanning 530004, China}

\begin{abstract}
POLAR-2 is a next-generation space astronomy platform led by China, with its core scientific objective focused on high-precision polarization measurements of gamma-ray bursts. As one of its key payloads, the Low-energy Polarization Detector (LPD) is designed to perform wide-field surveys to capture X-ray polarization information from gamma-ray bursts in the 2$\sim$10 keV energy range. This paper presents Topmetal-L, a dedicated charge-sensitive pixel sensor developed for the LPD prototype upgrade. Fabricated in a 130 nm CMOS process in 2024, the chip integrates a 356 $\times$ 512 pixel array with a pixel pitch of 45 $\mu$m. Each pixel incorporates a 26 $\times$ 26 $\mu$m$^{2}$ charge-collecting window and is capable of simultaneously outputting both energy and position information of deposited charges. Topmetal-L has been systematically optimized for power consumption, noise performance, and readout efficiency. It exhibits an input dynamic range of 0$\sim$4 ke$^{-}$, a typical charge-to-voltage conversion gain of 76.04 $\mu$V/e$^{-}$, an average equivalent noise charge of approximately 22.8 e$^{-}$, a sensitive area of 3.69 cm$^{2}$, and a total power consumption of 720 mW per chip. To meet the requirements of large-area, high-frame-rate readout for gas-based polarization detectors, a sentinel readout scheme is proposed, reducing the full-frame readout time to 730 $\mu$s. A prototype Topmetal-L-based gas polarization detection system was evaluated across key energies: it exhibited a residual modulation of 0.26\% $\pm$ 0.45\% at 5.90 keV, a modulation factor of 66.67\% $\pm$ 0.45\% for a linearly polarized 8.05 keV source, and a count rate saturated at 15 k counts$\cdot\text{cm}^{-2}\cdot\text{s}^{-1}$ when tested at 5.40 keV.

\end{abstract}

\keywords{CSA, Pixel sensor, Topmetal, X-ray polarization, Low noise}

\maketitle

\section{Introduction}

X-ray polarimetry, a pivotal subfield of astrophysics, has evolved through a symbiotic relationship between technological innovation and scientific discovery. Since the inaugural detection of the Crab Nebula in 1971, the discipline has progressed from confirmatory observations by the OSO-8 satellite\cite{01weisskopf1978precision}, through the monitoring of pulsar polarization evolution by Tsinghua's PolarLight CubeSat\cite{02feng2020re}, to the systematic and precise measurements enabled by the Imaging X-ray Polarimetry Explorer (IXPE) in 2021\cite{03weisskopf2022imaging,04xie2022vela}. While IXPE—utilizing grazing-incidence optics and large focusing mirror assemblies—offers high sensitivity, its limited agility hinders effective observations of transient sources. Its polarimetric analysis of the GRB 221009A afterglow, which yielded an upper limit of 13.8\% on the polarization degree in the 2$\sim$8 keV band\cite{05negro2023ixpe}, underscores the challenges faced by current instrumentation in capturing rapidly evolving phenomena such as gamma-ray bursts (GRBs)\cite{06de2018gamma}.\par
Unlocking the secrets of GRBs holds immense scientific value precisely because of the profound observational challenges they present. As one of the most extreme explosive phenomena in the universe, GRBs offer a unique laboratory for studying high-energy astrophysical processes. X-ray polarimetry has become a vital tool for probing extreme physics. Its application to the prompt emission of GRBs, which carries pristine information about the central engine, is key to unraveling radiation mechanisms, engine properties, and jet morphology and structure of magnetic field, thereby illuminating stellar evolution, black hole formation, and relativistic jet dynamics\cite{07tuo2024polarization,08zhong2025polarization,09mai2025time,10zalamea2011neutrino,11zhang2002analysis}. Nevertheless, polarimetric measurements of GRBs remain exceptionally difficult due to their transient nature and broad spectral energy distribution. Consequently, polarization data in the soft X-ray band are still notably scarce, underscoring both the need and the potential for future dedicated observations\cite{12toma2009statistical,13lundman2014polarization,14gill2020linear}.\par
The POLAR detector\cite{15produit2018design}, launched in 2016, marked a significant breakthrough by providing statistically meaningful polarization constraints for 14 out of the 55 Gamma-Ray Bursts it observed. These measurements not only validated the feasibility of space-borne polarimetry but also indicated a relatively low average polarization degree\cite{16zhang2019detailed}. This phenomenon may be linked to rapid polarization angle variations induced by the evolution of ultra-relativistic jets. To further investigate these findings and to deepen our understanding of key physical aspects such as GRB radiation mechanisms, jet structures, and magnetic field configurations, a collaborative Chinese-European team has proposed the next-generation instrument, POLAR-2\cite{17de2021development}. The primary scientific objective of POLAR-2 is to conduct high-precision polarimetry of GRBs, with a strategic extension of its energy coverage down to the soft X-ray range. A key component of POLAR-2, the Low-energy Polarization Detector (LPD)\cite{18feng2023orbit,19yi2024effectiveness,20wang2024low,21yi2024star,22feng2025polarization}, leverages its wide field-of-view survey capability in the 2–10 keV band to systematically capture polarization signals from the prompt emission and early soft X-ray flares of GRBs.\par
Achieving these scientific objectives presents formidable technical challenges for the LPD's core detection unit. It must integrate a large sensitive area for high detection efficiency, high spatial resolution for accurate reconstruction of the X-ray polarization direction, and low-noise characteristics for detecting faint signals—all within stringent power constraints and limited payload capacity. These inherently competing performance metrics pose a significant challenge to conventional polarimetry techniques. For instance, indirect measurement schemes based on Bragg diffraction or Thomson scattering\cite{23silver1990bragg,24shen1993complete}, though employed in early polarimetry, suffer from inherent limitations such as low efficiency, narrow energy bandwidth, or complex mechanical structures. While traditional CCDs offer imaging capabilities\cite{25hayashida1999optimization}, their inability to resolve the very short photoelectron tracks generated by soft X-rays prevents effective reconstruction of the polarization direction. These shortcomings make them unsuitable for meeting the LPD's requirements for a wide field-of-view, high dynamic response, and a compact payload.\par
In recent years, photoelectron track reconstruction based on the gas photoelectric effect has emerged as a promising solution to the limitations of conventional polarimetry. Its technical maturity and feasibility have been convincingly demonstrated by the successful in-orbit operations of PolarLight and IXPE, both of which utilize a Gas Pixel Detector (GPD)\cite{26bellazzini2006gas} structure based on  Gas Electron Multiplier (GEM)\cite{27sauli2016gas} and the XPOL-I\cite{28bellazzini2006direct}. Building upon the validated GPD, we have developed an optimized architecture for wide-field observation requirements: the Gas Microchannel Pixel Detector (GMPD)\cite{29feng2023charging,30yi2024pixel,31feng2024gas,32feng2024position}. This architecture synergistically integrates a Gas Microchannel Plate (GMCP)\cite{33feng2023spectral} with a custom charge-sensitive pixel chip, forming the basis for the LPD's detection unit. Within this architecture, large-array, low-power, charge-sensitive pixel chips play a central role. These chips integrate hundreds of thousands of miniature amplifiers on a single die, forming an imaging plane capable of parallel signal acquisition and processing. Their high spatial resolution is sufficient to accurately record the photoelectron tracks induced by X-ray photons in the gas detector, which enables the retrieval of the polarization information of the incident X-rays. Simultaneously, their low-noise and low-power characteristics directly determine the detector's sensitivity and long-term operational stability in the space environment.\par
In this context, our research focuses on the Topmetal-L, a large-array pixel chip custom-designed for the space-borne soft X-ray polarization detector. Topmetal-L is engineered to simultaneously meet the multiple demands of space applications, specifically high reliability, low power consumption, high readout speed, and two-dimensional imaging capability. It not only provides the fundamental hardware support for the LPD to achieve wide-field, high-precision polarimetry but also, owing to its rapid readout capability, represents a potent technological solution for capturing signals from transient sources such as GRBs. Following a brief introduction to the GMPD operating principle and LPD's pixel chip requirements in Section 2, the paper details the design and implementation of Topmetal-L in Section 3. Section 4 characterizes its electrical properties. Section 5 then assesses its capabilities for X-ray polarimetry when deployed within a gas detector. The paper culminates in Section 6 with a conclusion and a discussion of future prospects.\par

\section{Detector Principles and LPD Requirements}

\begin{figure}[htbp]
    \centering
    \includegraphics[width=0.48\textwidth]{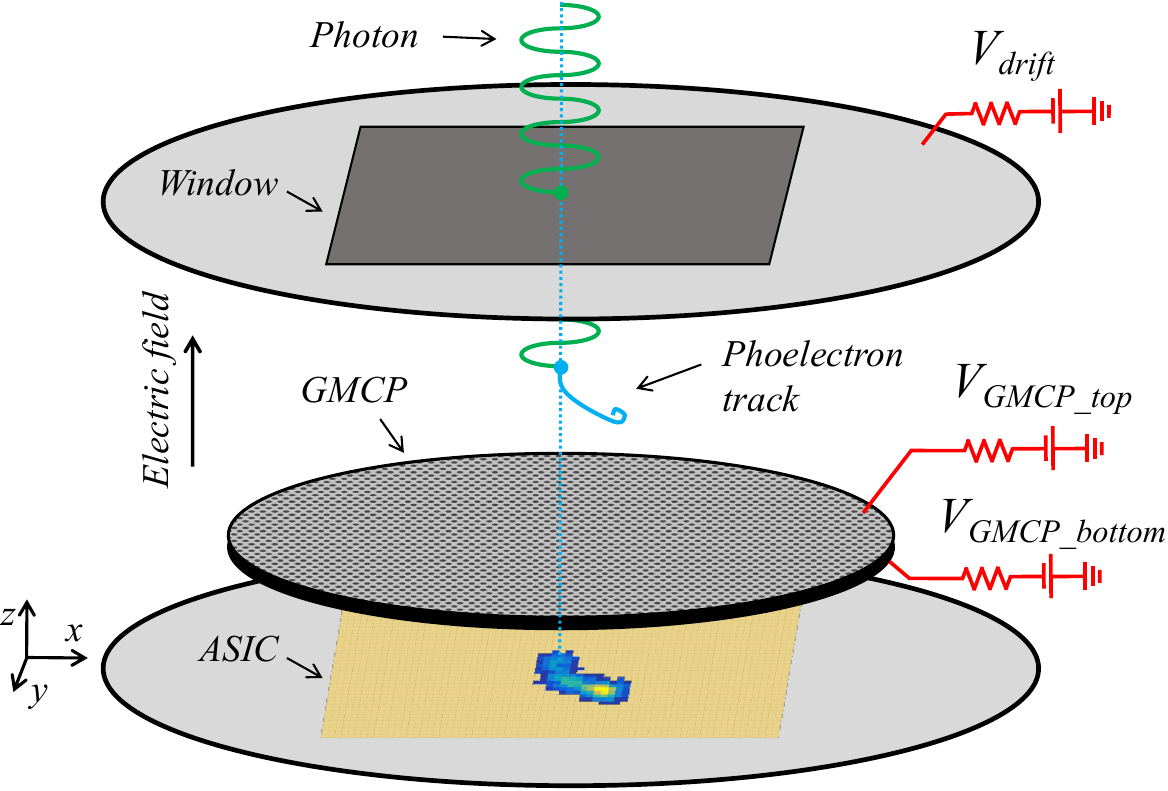}
    \caption{Schematic diagram of the GMPD detection principle.}
    \label{fig:1}
\end{figure}

The core components of the GMPD include a GMCP and a Topmetal pixel sensor. Both the top and bottom surfaces of the GMCP are metallized to form electrodes; applying a voltage potential between these electrodes creates a strong electric field within its microchannels, essential for electron multiplication. During operation, these two core components are integrated into a sealed chamber filled with a specific working gas. As illustrated in Fig.~\ref{fig:1}, a planar electrode with a thin beryllium window is positioned at the top of the chamber to transmit incident X-rays. The Topmetal pixel chip, serving as the anode, is precision-aligned beneath the GMCP, with its charge-collection surface facing upward and parallel to the GMCP's bottom surface. Within this configuration, a specific sequence of negative bias voltages applied to the respective electrodes establishes an electric field throughout the chamber volume, driving the directional drift of electrons.\par
The complete detection process begins with an incident X-ray photon. After passing through the beryllium window, the photon is absorbed by a gas atom via the photoelectric effect, ejecting a photoelectron whose initial direction of motion is modulated by the polarization state of the incident photon. As this photoelectron travels through the gas, it generates a substantial number of primary electronion pairs along its trajectory via ionizing collisions. These primary electrons drift towards the GMCP, guided by the electric field. Upon entering the high-field region within the microchannels, they undergo avalanche multiplication, producing a cloud of secondary electrons amplified by approximately four orders of magnitude. The avalanche-induced electrons continue to drift downward, where a small fraction is absorbed by the GMCP's bottom electrode, and the vast majority are ultimately collected by the pixel chip. The charge collected on the GMCP's bottom is measured by a dedicated circuit in the GMPD for accurate energy readout\cite{34fan2023front}. The charge signal collected by the electrode is converted into a voltage pulse by the charge-sensitive amplifier (CSA) integrated in each pixel for readout. The system outputs information for each event, including the address of every fired pixel and its corresponding charge information. These signals are then aligned using temporal coincidence. The polarization characteristics of the incident X-rays are subsequently reconstructed through morphological analysis and statistical fitting of the two-dimensional track images accumulated from a large number of individual X-ray events\cite{35huang2021simulation,36zhang2022simulation}.\par
The first-generation LPD detector prototype, which utilized the Topmetal-II$^{-}$ sensor\cite{37li2021preliminary}, successfully validated the feasibility of the synergistic GMCP and Topmetal architecture. This prototype, developed into the CXPD\cite{41liu2025wide,38wang2023electronics} instrument, has since demonstrated successful on-orbit operation. This chip incorporates a 72 $\times$ 72 pixel array, with an effective charge-sensitive area of 6 $\times$ 6 mm$^2$ and a pixel pitch of 83 $\mu$m. The top-metal electrode of each individual pixel measures a 15 $\times$ 15 $\mu$m$^2$ area is directly exposed for charge collection. Each pixel electrode is connected to the input node of its corresponding CSA, achieving an Equivalent Noise Charge (ENC) of 13.9 e$^{-}$ at room temperature. It employs a Rolling Shutter readout scheme with a frame rate of 386 Hz, and the entire chip consumes about 1 W of power. The GMCP effectively dissipates charge accumulated within its microchannels via its bulk resistance, thereby maintaining gain stability across different operational durations and count rates. The bulk material of the GMCP consists of low-outgassing materials, and when combined with a low-outgassing rate housing and laser welding sealing technology, this further ensures the long-term stability of the detector. In performance characterization, the prototype demonstrated a modulation factor of 41.28\% $\pm$ 0.64\% for a 4.51 keV polarized X-ray beam, and a residual modulation of 1.96\% $\pm$ 0.58\% was observed across the entire sensitive area when illuminated with a 5.90 keV unpolarized source. These results indicate the superior performance of the GMCP for gas-based polarimetry detection, while the Topmetal-II$^{-}$, owing to its low-noise characteristics, enabled high-quality photoelectron track imaging.\par
However, the stringent resource constraints of the space station and the scientific objectives of the LPD's wide-field-of-view GRB monitoring survey mean that the limitations of the Topmetal-II$^{-}$ in terms of power consumption, sensitive area, and readout speed significantly hinder its direct applicability for LPD. As outlined in Table~\ref{tab:1}, the LPD technical specifications impose more demanding requirements regarding effective area, power consumption per-detection-unit, and count rate capability. A simple tiling approach using multiple Topmetal-II$^{-}$ to increase the sensitive area presents several critical issues. Although the frame rate of a single chip might meet the count rate requirement, ensuring the total count rate for a single detection unit necessitates a multi-channel parallel readout architecture. This would substantially increase the power consumption of the detection unit, as well as the complexity and total power dissipation of the peripheral electronics. Furthermore, due to the chip's relatively low fill factor (ratio of sensitive to total area), achieving the target effective area of 3.6 cm$^2$ for a single detection unit would require integrating approximately ten chips, resulting in poor area utilization and significant resource waste, which is unacceptable in spaceborne payload design. Multi-chip tiling also introduces additional design challenges, posing significant engineering hurdles related to maintaining internal electric field homogeneity and managing on-orbit thermal dissipation.\par

\begin{table}[htbp]
\centering
\caption{LPD Technical Design Indicators}
\label{tab:1}
\begin{tabular}{ll}
\toprule
\textbf{Technical Parameter} & \textbf{Specification} \\ 
\midrule
Detector matching & 9 detection units \\[0.5ex]
Energy range & 2$\sim$10 keV \\[0.8ex]
Effective area per detection unit & \(\geq 3.6 \, \text{cm}^2\) \\[0.8ex]
Total effective area & \(\geq 30 \, \text{cm}^2\) \\[0.8ex]
Power consumption per detection unit & \(\leq 0.9 \, \text{W}\) \\[0.8ex]
Detector module power consumption & \multirow{2}{*}{\(\leq 25 \, \text{W}\)} \\
\quad (including 3 detection units) & \\[0.8ex]
Energy resolution & \(\leq 25\%\,@\,5.90 \, \text{keV}\) \\[0.8ex]
Modulation factor & \(\geq 40\%\,@\,5.90 \, \text{keV}\) \\[0.8ex]
Time resolution & \(\leq 3.3 \, \text{ms}\) \\[0.8ex]
Geometric field of view & \(90^\circ\) \\[0.8ex]
Count rate & \(\geq 350 \, \text{counts}\cdot\text{cm}^{-2}\cdot\text{s}^{-1}\) \\[0.8ex]
\bottomrule
\end{tabular}
\end{table}

\begin{figure}[htbp]
    \centering
    \includegraphics[width=0.48\textwidth]{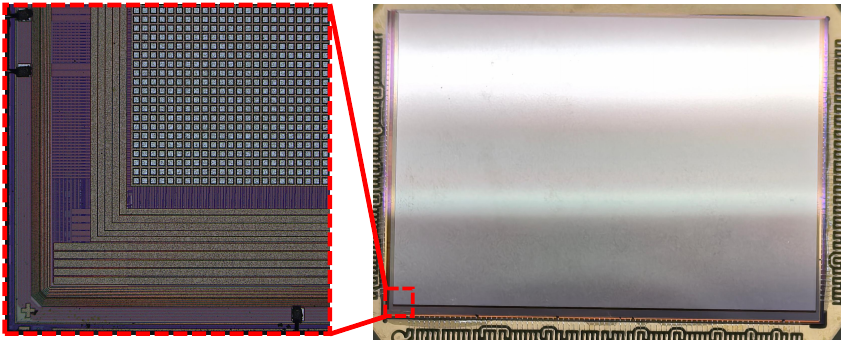}
    \caption{Optical photograph of Topmetal-L, showing the overall appearance and a detailed view of the pixel array.}
    \label{fig:2}
\end{figure}

To address the aforementioned challenges, the LPD urgently requires a charge-sensitive pixel readout chip featuring low noise, low power consumption, a large sensitive area, and high readout speed. Driven by this requirement, our research team has designed the Topmetal-L pixel sensor as an Application-Specific Integrated Circuit (ASIC) customized for the LPD. A photograph of the chip is presented in Fig.~\ref{fig:2}.

\section{Topmetal-L Implementation}

\begin{figure}[htbp]
    \centering
    \includegraphics[width=0.48\textwidth]{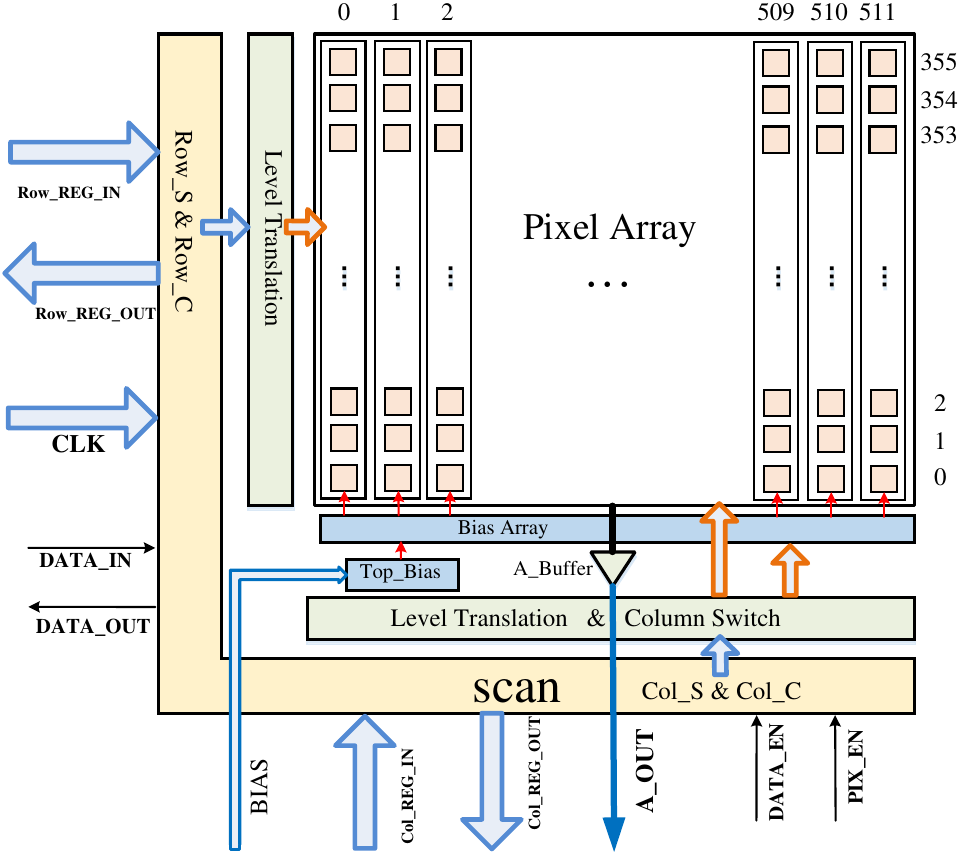}
    \caption{Overall architecture of the Topmetal-L.}
    \label{fig:3}
\end{figure}

The Topmetal-L was fabricated in a 130 nm CMOS process with four wells. Its I/O pins are strategically placed along the left, right, and bottom edges, a layout optimized to accommodate the mechanical structure and electrical interconnection requirements of the LPD detector. The chip measures 17 $\times$ 24 mm$^2$ in total and features a 356 (row) $\times$ 512 (column) pixel array. The pixel pitch is 45 $\mu$m, yielding an effective sensitive area of 16.02 $\times$ 23.04 mm$^2$. The overall architecture is illustrated in Fig.~\ref{fig:3}. For large pixel array designs requiring high count rate performance, partitioning the array into multiple output channels is an effective approach to ensure fast readout. However, given the need for track reconstruction and polarization analysis, inter-channel inconsistencies cannot be neglected in this application. When a track straddles the boundary between channels, significant systematic effects may be introduced, compromising the determination of the emission direction. Therefore, the Topmetal-L design adopts a structure with a single analog output channel shared across the entire array. A global biasing array generates reference voltages, which are distributed to the pixel array via column-level biasing circuits. The output of the pixel array is controlled by a peripheral scanning module, which can be configured via registers to define scanning regions and modes. \par

At the circuit design level, the chip uses a 1.2 V supply for the digital logic to achieve low power consumption, while the analog section employs a 2.0 V supply. Control signals between these domains are bridged by a level-shifter array with low quiescent power consumption. At the layout level, analog and digital modules are isolated using deep N-well technology, significantly reducing substrate-coupled digital switching noise interference with sensitive analog circuits. 
While achieving an substantial sensitive area, Topmetal-L maintains a total power consumption of approximately 720 mW, with 680 mW in the analog section and 40 mW in the digital section. This combination of a large sensitive area and relatively low power consumption fully meets the stringent technical requirements of the LPD detector.\par

\begin{figure}[htbp]
    \centering
    \subfigure[]{
        \includegraphics[width=0.235\textwidth]{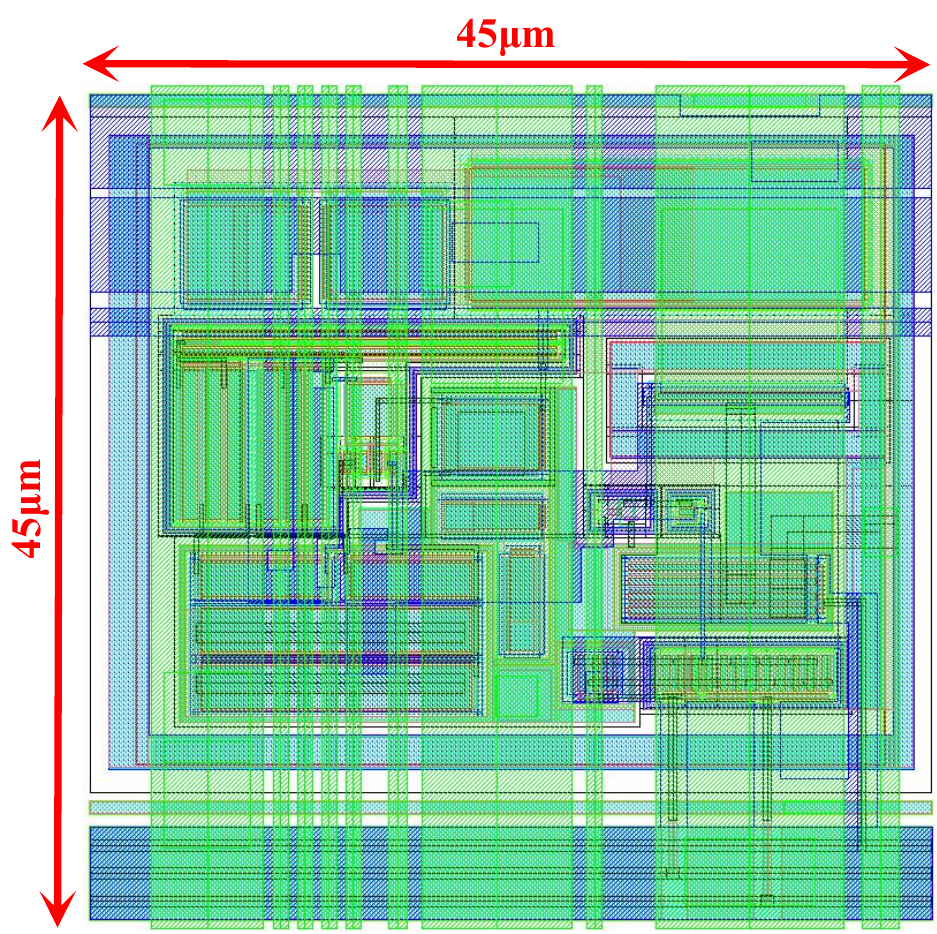}
        \label{fig:4(a)}
    }
    \hfill
    \subfigure[]{
        \includegraphics[width=0.225\textwidth]{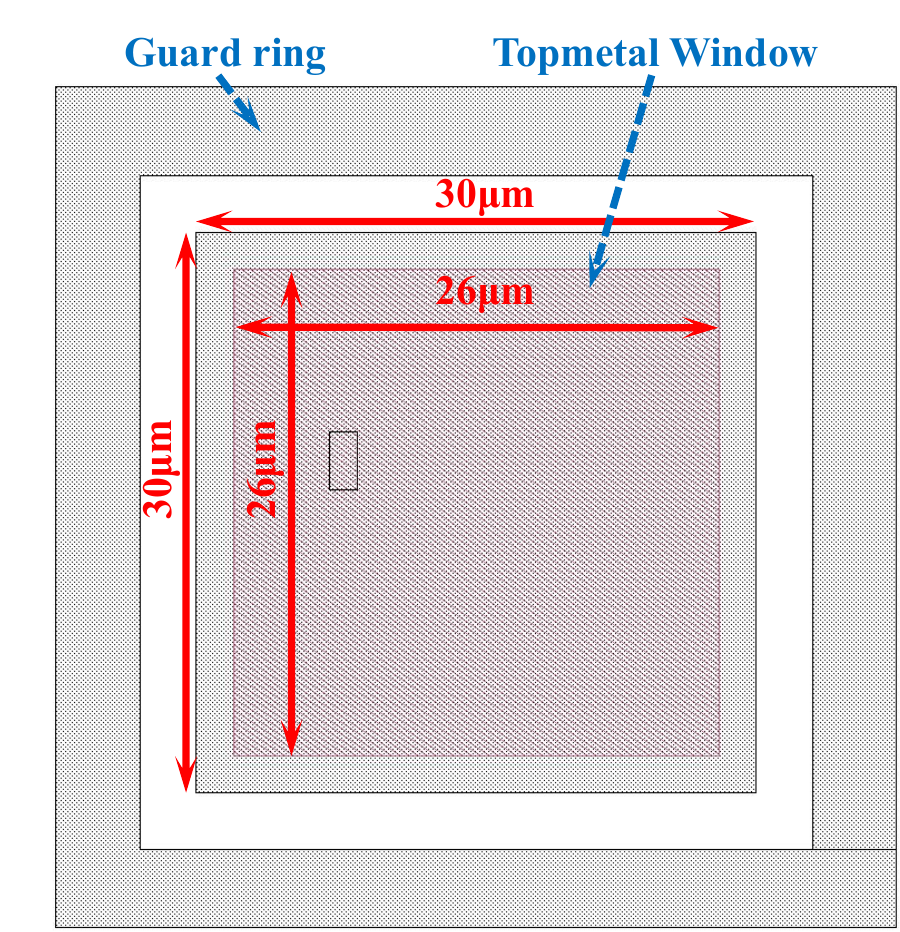}
        \label{fig:4(b)}
    }
    \caption{(a) Layout of a single pixel cell in the Topmetal-L; (b) Top-metal structure.}
\end{figure}

The layout of a single pixel unit in Topmetal-L is shown in Fig.~\ref{fig:4(a)}. The pixel cell dimensions are 45 $\times$ 45 $\mu$m$^2$. Fig.~\ref{fig:4(b)} shows the top-metal structure of a single pixel. In the GMPD application, Topmetal-L is designed to collect charge signals drifting down from the GMCP. The exposed top metal window at the pixel center functions as the charge-collecting electrode. The electrode measures 30 $\times$ 30 $\mu$m$^2$, with an exposed window size of 26 $\times$ 26 $\mu$m$^2$. Each electrode is surrounded by a guard ring within the same metal layer. Areas other than the exposed electrode are covered by a passivation layer. The spacing between the electrode and the guard ring is 2 $\mu$m, forming a capacitance $C_g$ of approximately 5.5 fF. The guard rings of all pixels are interconnected and eventually tied to a common pad. During pixel array calibration, a voltage step signal can be injected into this guard ring to simulate charge injection. When Topmetal-L is used for actual charge signal measurement, the guard ring must be grounded. \par

\begin{figure}[htbp]
    \centering
    \includegraphics[width=0.48\textwidth]{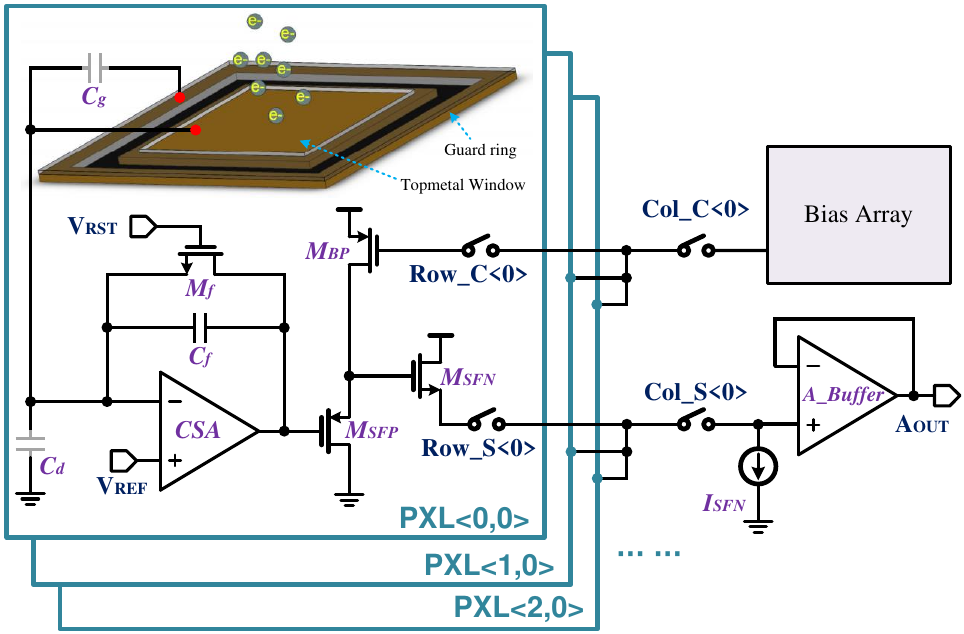}
    \caption{Architecture of the analog channel in Topmetal-L.}
    \label{fig:5}
\end{figure}

The architecture of the analog channel in Topmetal-L is illustrated in Fig.~\ref{fig:5}. The core functional modules of a single pixel include: a Topmetal collecting electrode, a CSA, a source follower as a buffer (SFP), an input transistor of the output-stage source follower, and two row control switches (Row\_C and Row\_S), which control the tail current biasing of the SFP and the readout enable of the analog signal, respectively. Peripherally, each column is equipped with two column-level control switches (Col\_C and Col\_S), managing column-level biasing and column-level output signals, respectively. All analog channels across the entire array share a single second-stage source follower and its tail current source, with the final output driven off-chip through a common analog buffer. By configuring the row and column control switches, the bias current of the SFP in the target pixel is first enabled to activate it. The corresponding Row\_S and Col\_S are then switched on to gate the pixel’s analog signal onto the output bus.\par

\begin{figure}[htbp]
    \centering
    \includegraphics[width=0.48\textwidth]{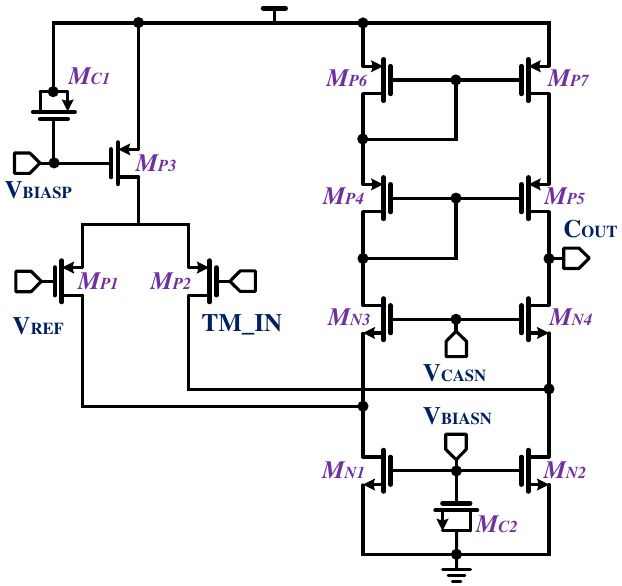}
    \caption{Operational amplifier architecture within the CSA.}
    \label{fig:6}
\end{figure}

The total operating current per pixel is strictly limited to 1.6 $\mu$A to meet the low-power target for the large array, which restricts the integration of excessive functional modules within a single pixel. Consequently, the design focuses primarily on the CSA, which is most critical for both power consumption and performance. The feedback network of the CSA consists of a designed metal–metal capacitor ($C_f$) of approximately 1 fF and an NMOS transistor ($M_f$). This design choice is based on two considerations: first, replacing a conventional large-area feedback resistor with a MOS transistor significantly saves layout space within the limited pixel area; second, the NMOS operates as a voltage-controlled resistor, allowing flexible adjustment of the CSA discharge time by tuning its gate voltage ($V_{RST}$), thereby accommodating different readout mode requirements. When the operational amplifier gain is sufficiently high, the charge gain of the CSA is given by Eq. (3.1):\par

\[
A_{v,Q} = \frac{V_{\text{out}}}{Q_m} \approx -\frac{1}{C_f}
\tag{3.1}
\]\par

where $V_{out}$ is the output voltage amplitude of the CSA and $Q_{in}$ is the input charge quantity.\par

The CSA employs a folded-cascode operational amplifier as its core. Its high gain ensures precision in charge measurement, while its favorable input and output swing meets dynamic range requirements under low-voltage supply. The specific circuit structure of this operational amplifier is shown in Fig.~\ref{fig:6}. As the front end of the entire readout chain, the CSA is the primary contributor to circuit noise, with its operational amplifier being the dominant noise source. The noise contributions of the current-mirror transistors are approximately equal, and the contribution of each pair of MOS transistors to the equivalent input noise voltage is given by Eq. (3.2):\par

\begin{equation}
\begin{split}
\overline{v_{n,in,CG}^{2}} \approx & 8kT \gamma \biggl(1+\frac{g_{m,N1}}{g_{m,P1}} + \frac{g_{m,P3}}{g_{m,P1}^2}\biggr) + 4kT \gamma \frac{g_{m,P6}}{g_{m,P1}} \\
& + \frac{2K_P}{C_{\text{OX}}(WL)_{P1}} \frac{1}{f}
\end{split}
\tag{3.2}
\end{equation}

where $K_P$ is the flicker noise coefficient of the PMOS, $C_{OX}$ is the gate oxide capacitance per unit area, $(WL)_{P1}$ represents the channel area of $M_{P1}$, $k$ is Boltzmann’s constant, $T$ is the absolute temperature, and $\gamma$ is the noise coefficient. The first term represents the thermal noise contributions from the input differential pair and the two sets of current-mirror loads, while the second term corresponds to the thermal noise contribution of the tail current source to the equivalent input noise. This analysis suggests that the transconductance of the input transistors should be maximized, whereas the transconductance values of the current mirrors and the current source should be minimized, thereby suppressing the contributions of these devices to the equivalent input noise. The third term represents the flicker noise contribution of the input pair. In conventional silicon CMOS processes, the probability of PMOS carriers (holes) being trapped by interface states is much lower than that of NMOS carriers (electrons), resulting in a smaller flicker noise coefficient for PMOS. Therefore, a PMOS is selected for the input pair. The flicker noise of only the input pair is considered because, when referred to the input, the flicker noise from other transistors is divided by the overall amplifier gain and becomes negligible. Additionally, capacitors $M_{C1}$ and $M_{C2}$ are introduced at the gate nodes of the current mirrors to suppress noise coupling from the bias bus. Eq. (3.2) does not discuss the contribution of the cascode transistors to the input noise because their contribution is essentially negligible. The contribution of the cascode transistors is given by Eq. (3.3):\par

\begin{equation}
\begin{split}
\overline{v_{n,in,AMP}^{2}} \approx & 8kT \gamma \biggl[\frac{1}{g_{m,N3}g_{m,P1}^2(r_{o,N1}//r_{o,P1})^2} \\
& + \frac{1}{g_{m,P4}g_{m,P1}^2r_{o,P6}^2}\biggr]
\end{split}
\tag{3.3}
\end{equation}

It can be seen that its contribution is severely attenuated.\par
For the overall noise of the CSA, the output noise under no input condition can be used for representation, as shown in Eq. (3.4):\par

\begin{equation}
\begin{split}
\overline{v_{n,in,AMP}^{2}} \approx & \overline{v_{n,in,AMP}^{2}} \biggl(\frac{C_d+C_f}{C_f}\biggr)^2+\biggl[2qI_{g,P1}+\\
&4kT\gamma g_{m,f}  + \frac{K_N}{C_{\text{OX}}(WL)_{f}}\biggr]\left|\frac{r_{o,f}}{1+sr_{o,f}C_f}\right|^2
\end{split}
\tag{3.4}
\end{equation}

where $I_{g,P1}$ denotes the gate leakage current of the input transistor, $g_{m,f}$ and $r_{o,f}$ represent the transconductance and output resistance of $M_f$. This considers the shot noise of the input transistor and the contribution of transistor $M_f$ to the output noise. The equivalent input noise of the CSA is still dominated by the input transistor of the amplifier. However, at low frequencies, the noise contribution of $M_f$ still cannot be neglected.It is noteworthy that the detector capacitance $C_d$ at the input directly influences the equivalent input noise. In the application of Topmetal-L in a GMPD, the so-called detector capacitance physically corresponds to the parallel-plate capacitance formed between the bottom electrode of the GMCP and the top-metal electrode of the chip. Since the electrode size is limited and the plate separation is on the order of millimeters, this capacitance is almost negligible. Therefore, the internal parasitic capacitances of the chip (primarily including the input transistor gate capacitance, metal interconnect capacitance, and protection structure capacitance) actually constitute the main part of $C_d$. This characteristic imposes stricter requirements on circuit and layout design. The design process requires trade-offs among multiple competing factors: for instance, the dimensions of the input and feedback transistors must meet transconductance and noise targets while considering area and power consumption; The size of the top collecting electrode should not be too small to maintain charge collection efficiency, while an excessively large size would introduce greater parasitic capacitance; while transistor dimensions cannot be indiscriminately reduced, lest their flicker noise coefficient increases. The final design is the result of systematically seeking a balance among noise performance, collection efficiency, power consumption, and area.\par
In the readout architecture design, Topmetal-L adopts a single analog output channel scheme to balance overall chip power consumption, peripheral circuit complexity, and track reconstruction quality. However, this approach also introduces new challenges. Due to the large chip area, pixels located at the edges of the array must connect to the common output bus and analog buffer via long metal lines. Shared bus structure introduces significant parasitic capacitance and resistance, severely limiting the output channel bandwidth. To ensure the quality of the sampled signal, the scanning clock frequency is limited to 10 MHz. If a conventional rolling shutter readout were used, completing a full frame would take approximately 18.3 ms—a frame rate far below the detection requirements of the LPD for gamma-ray bursts' high count rates. To address the localized spatial distribution characteristic of photoelectron tracks in gas polarization detection, this study proposes an efficient triggered scanning method named sentinel readout to significantly enhance the effective readout rate. This mode operates in two distinct phases:\par

1. Phase 1 is the Awaiting Trigger phase, where the system performs a low-duty-cycle, interval scan across the entire pixel array. The scan starting point and step size are configured via registers. When the step size is greater than 1, the row or column address pointer skips a specified number of pixels after each readout, thus sampling only the uniformly distributed sentinel pixels. The step size is optimized based on the typical spatial dimensions of the target photoelectron tracks.\par
2. Once the signal from any sentinel pixel exceeds a preset threshold, the system immediately switches to Phase 2, the region scanning mode. A local region is automatically defined, centered on the triggered sentinel pixel, with dimensions based on the step size, and is scanned continuously. To ensure complete capture of tracks that may span multiple sentinel regions, the areas corresponding to the adjacent sentinels surrounding the triggered one are typically also included in the readout. After the region scanning is complete, the system automatically resets and returns to the sentinel scanning state of Phase 1.\par

The entire readout process requires coordination with an external Analog-to-Digital Converter (ADC) and a Field-Programmable Gate Array (FPGA) for implementation. Detailed control procedures and implementation can be found in \cite{42Zhou2025wide}. The sentinel readout mode will undergo system-level performance validation in subsequent sections.\par

\section{Electrical Characterization}

\begin{figure}[htbp]
    \centering
    \includegraphics[width=0.42\textwidth]{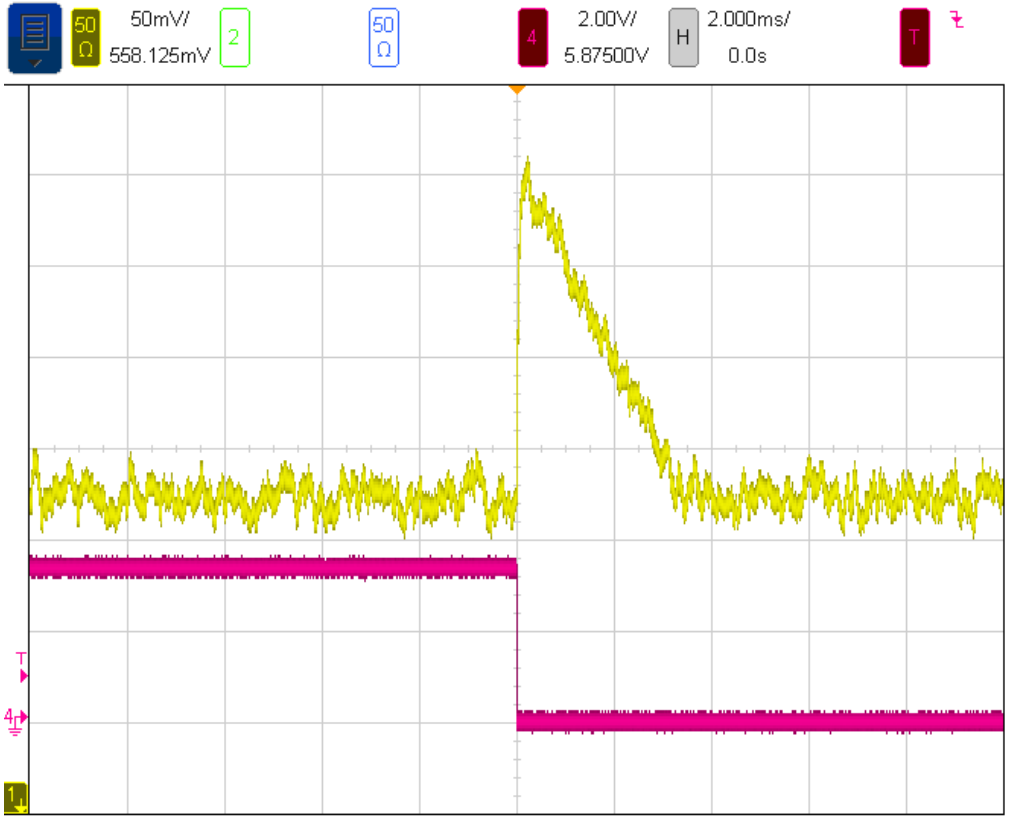}
    \caption{Single-pixel analog output signal waveform (yellow) following the injection of a negative step signal into the guard ring, with the red trace representing the synchronous trigger signal for the injection.}
    \label{fig:7}
\end{figure}

To evaluate the electrical performance of Topmetal-L, comprehensive chip-level tests were conducted at room temperature and under atmospheric conditions. For single-pixel characterization, the chip readout was fixed to a specified pixel through register configuration, and a negative step signal was injected into the guard ring using a signal generator to simulate charge injection into the top-metal electrode. The amount of injected charge is determined by Eq. (4.1):\par
\[
Q_{in} = V_gC_g \tag{4.1}
\]\par
where $V_g$ is the step voltage change. During testing, $V_{REF}$ was set to 410 mV and $V_{RST}$ to 580 mV, with all tests performed using this bias configuration. The synchronization signal of the input negative step signal and the analog output waveform of a single pixel are shown in Fig.~\ref{fig:7}. The negative step voltage provided by the signal generator was $V_g$ = 60 mV. The CSA output is a pulse signal with a decay time of about 3.2 ms.\par

\begin{figure}[htbp]
    \centering
    \includegraphics[width=0.48\textwidth]{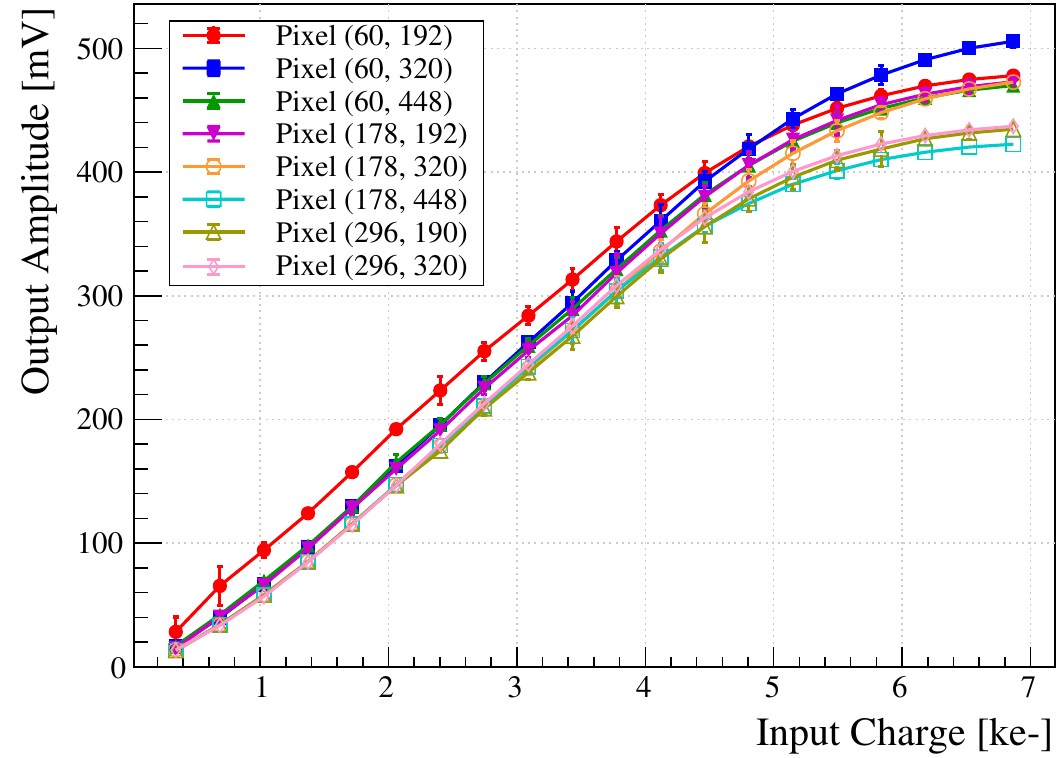}
    \caption{Input–output response curves of eight different pixels.}
    \label{fig:8}
\end{figure}

Fig.~\ref{fig:8} displays the input-output response curves of eight different pixels. Within the input range of 0$\sim$4 ke$^{-}$, the corresponding output linear range is approximately 0$\sim$350 mV. All pixels exhibit acceptable linear response to the input charge, with a typical integral nonlinearity (INL) of 1.78\%. To meet low-power design requirements, the analog supply voltage was appropriately reduced while ensuring no impact on track reconstruction, though this limited the amplifier's output swing. Additionally, the cascaded structure of the two-stage source follower also constrained the output range to some extent. The curves reveal variations in gain characteristics and dynamic range among pixels at different locations, primarily attributable to non-uniform distribution of parasitic resistance from pixels to the output and inherent device mismatch effects.\par

\begin{figure}[htbp]
    \centering
    \includegraphics[width=0.48\textwidth]{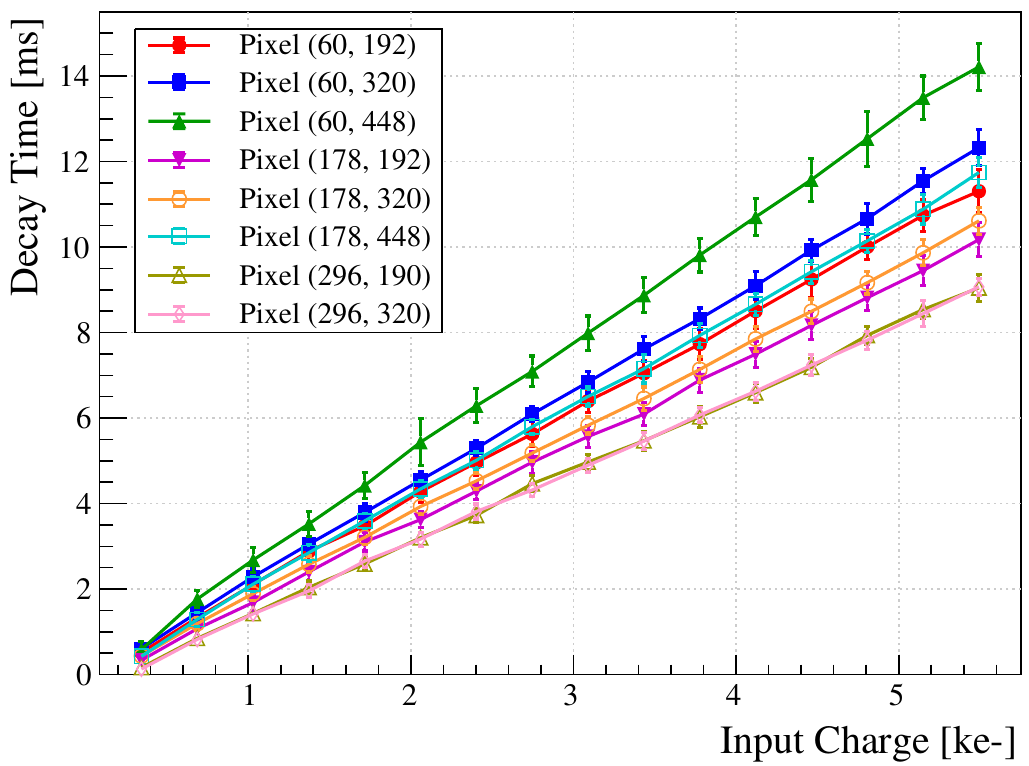}
    \caption{Decay time versus input signal for eight different pixels.}
    \label{fig:9}
\end{figure}

An interesting observation from Fig.~\ref{fig:7} is that the output recovery of a traditional resistive- capacitive feedback CSA to baseline generally follows an exponential decay process, whose characteristics are determined by the time constant:\par
\[
\tau = R_fC_f \tag{4.2}
\]\par

where $R_f$ represents the feedback resistance. In contrast, the structure employed in this design, which utilizes an NMOS transistor to replace the feedback resistor, exhibits an approximately linear decay process. We analyzed the relationship between decay time and input signal for eight different pixels, as shown in Fig.~\ref{fig:9}. The results indicate that this relationship also demonstrates favorable linearity, with a typical INL of 1.04\%. This characteristic is determined by its operational principle. In large-scale array design, although the sentinel readout mode significantly enhances readout speed, the overall readout time still needs to approach the millisecond level.\par

To ensure as complete collection of track energy as possible, the CSA is required to have a long decay time. For a feedback capacitance of 1 fF, achieving a time constant close to the millisecond level would require a feedback resistance on the order of $10^{12}$ $\Omega$. Implementing such a large resistance within a pixel is entirely impractical. Therefore, we employ an NMOS transistor operating in the deep sub-threshold region to achieve this high-resistance characteristic. The current expression for an NMOS operating in the deep sub-threshold region is given by Eq. (4.3):\par

\begin{equation}
\begin{split}
I_{D} = I_{D0}\frac{W}{L} e^{\frac{V_{GS}-V_{TH}}{nV_T}}(1-e^{-\frac{V_{DS}}{nV_T}})
\end{split}
\tag{4.3}
\end{equation}

where $I_{D0}$ is a process-dependent parameter, $n$ is the subthreshold slope factor, and $V_T = kT/q$ is the thermal voltage, with a typical value of approximately 26 mV at room temperature. When $V_{DS}$ exceeds 78 mV, the exponential term related to $V_{DS}$ can be approximately neglected. This condition is generally satisfied during the main phase of the CSA output response, allowing the feedback NMOS to be equivalently modeled as a constant current source. Consequently, the discharge process of the feedback capacitor $C_f$ is dominated by this constant current, resulting in a linear recovery slope in the output waveform. Although the discharge current gradually decreases as $V_{DS}$ drops to a lower level toward the end of the decay process, causing deviation from ideal linear behavior, this nonlinear phase constitutes an extremely small portion of the overall signal. Furthermore, due to circuit noise, such deviations are typically not significant in experimental observations. Under this linear decay mechanism, the traditional description of the decay process using the time constant $\tau$ is no longer applicable.\par
It is more reasonable to characterize using the discharge rate, which can be expressed as:\par

\[
v_{Decay} = \frac{dV}{dt}=\frac{I_D}{C_f} \tag{4.4}
\]\par

Based on the measured data, the typical value of this discharge speed is approximately 37.28 mV/ms. The pixel-to-pixel consistency of the decay time versus input signal curves shown in Fig.~\ref{fig:9} is noticeably lower than that of the gain curves. This is primarily attributed to the higher sensitivity of the MOSFET transconductance to process mismatch when operating in the deep sub-threshold region. This phenomenon represents a key challenge in ensuring pixel uniformity for low-power, pixel-level chip design. To mitigate this issue, a per-pixel calibration strategy for each individual chip can be employed, although it would significantly increase the testing and calibration workload.\par

\begin{figure}[htbp]
    \centering
    \includegraphics[width=0.48\textwidth]{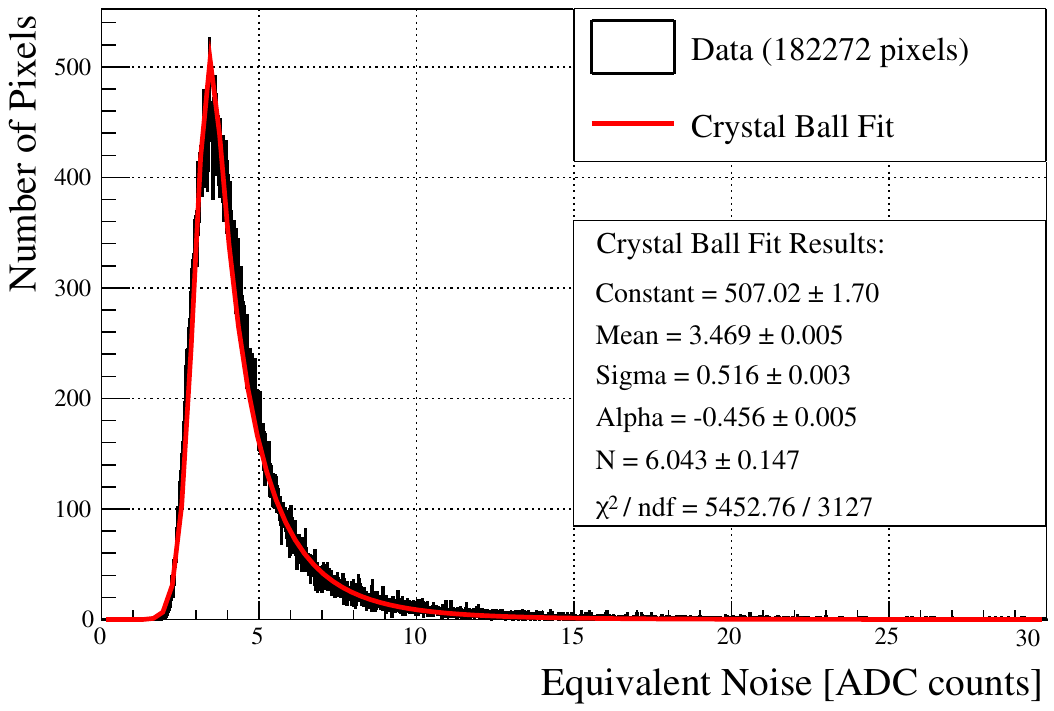}
    \caption{Noise distribution across all pixels. The horizontal axis is given in ADC counts, with a conversion factor of approximately 0.5 mV/ADC count.}
    \label{fig:10}
\end{figure}

When analyzing the noise performance of Topmetal-L, the guard ring was first grounded, and the entire array was scanned. A 12-bit ADC was used for sampling to acquire baseline data for each pixel. The noise voltage of each pixel was obtained by calculating the root mean square (RMS) of its baseline data. Fig.~\ref{fig:10} shows the noise statistics for all 182,722 pixels. The average noise, fitted using a Crystal Ball function, is approximately 3.47 ADC counts. The conversion factor to voltage is about 0.5 mV/ADC count. Subsequently, an 80 mV negative step signal was applied to the guard ring. A uniform selection of 1000 pixels was used to statistically determine a typical gain value of 76.04 $\mu$V/e$^{-}$. In nuclear electronics, ENC is commonly used to characterize the noise performance of a system, defined as:
\[
ENC = \frac{V_{n,out,rms}}{S}=\frac{V_{n,out,rms}}{A_{OUT}} V_gC_g\tag{4.5}
\]

where $S = A_{OUT}/Q_{in}$ is the charge sensitivity of the system, and $V_{n,out,rms}$ is the RMS of the measured output noise voltage. For an 80 mV negative step input signal, the equivalent injected charge is 2.747 ke$^{-}$. Under this condition, the calculated ENC is approximately 22.8 e$^{-}$, the value includes the noise contributions from the Topmetal-L and the external front-end electronics. This noise performance shows a certain gap compared to the 13.9 e$^{-}$ achieved by Topmetal-II$^{-}$. This result stems from a critical trade-off in the pixel structure design: to enhance charge collection efficiency, Topmetal-L significantly increases the electrode area within each pixel for receiving charge signals. The window area threefold, from 225 $\mu$m$^{2}$ in Topmetal-II$^{-}$ to 676 $\mu$m$^{2}$. This larger area provides higher charge collection efficiency, and the resulting signal gain outweighs the increased noise, leading to a net improvement in the signal-to-noise ratio. The observed noise level is thus a justified trade-off.

\section{System-Level Validation}

\begin{figure}[htbp]
    
    \centering
    \includegraphics[width=0.48\textwidth]{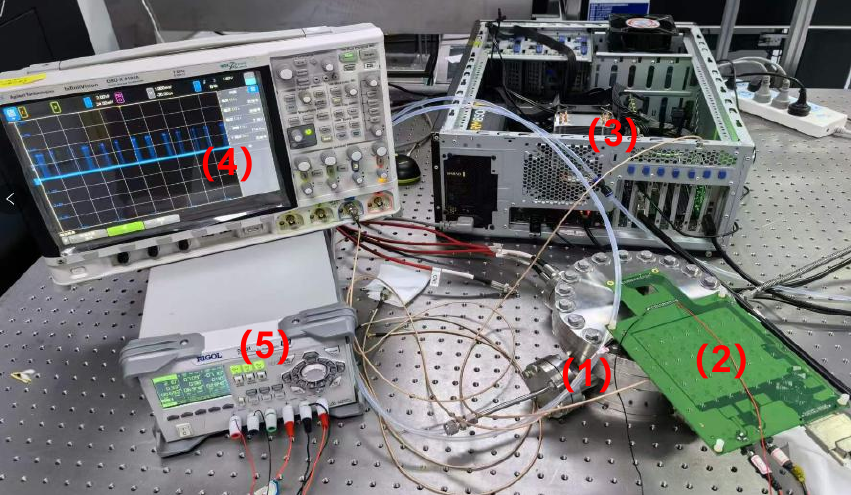}
    \caption{Schematic of the soft X-ray test platform for Topmetal-L. (1) Gas-flow chamber: filled with specific working gas and integrated with GMCP and Topmetal-L; (2) Front-end readout board: provides bias voltages and receives analog signals from the chip; (3) Server host: connected to the back-end data acquisition card via PCIe interface for high-speed data transfer and real-time analysis; (4) Oscilloscope: monitors the output waveform and operational status of the chip; (5) DC power supply: powers the electronics system.}
    \label{fig:11}
\end{figure}

To validate the feasibility of Topmetal-L for GMPD applications, a soft X-ray test platform was established under room temperature and atmospheric pressure. As illustrated in Fig.~\ref{fig:11}, the platform mainly consists of a gas-flow chamber, a front-end board, a server host connected via a PCIe (Peripheral Component Interconnect Express) interface to a high-speed data acquisition card, an oscilloscope, a DC power supply, and a high-voltage module (not shown in the figure). The GMCP and Topmetal-L are integrated inside the gas chamber according to the structure shown in Fig.~\ref{fig:1}, with a cathode to GMCP upper surface distance of 10 mm and a GMCP bottom surface to Topmetal-L distance of 3 mm. The high-voltage module supplies the required bias voltages: cathode voltage $V_{drift}$ = -3600 V, GMCP upper surface voltage $V_{GMCP\_top}$ = -1650 V, and GMCP lower surface voltage $V_{GMCP\_bottom}$ = -500 V. The GMCP used has a thickness of 300 $\mu$m, with pore diameter and pitch of 50 $\mu$m and 60 $\mu$m, respectively, and a bulk resistance of 2 G$\Omega$. The chamber is filled with a gas mixture of 40\% helium (He) and 60\% dimethyl ether (DME, C$_2$H$_6$O). A $^{55}$Fe source is used, emitting 5.90 keV characteristic X-rays incident from above the cathode.

\begin{figure}[htbp]
    
    \centering
    \subfigure[]{
        \includegraphics[width=0.23\textwidth]{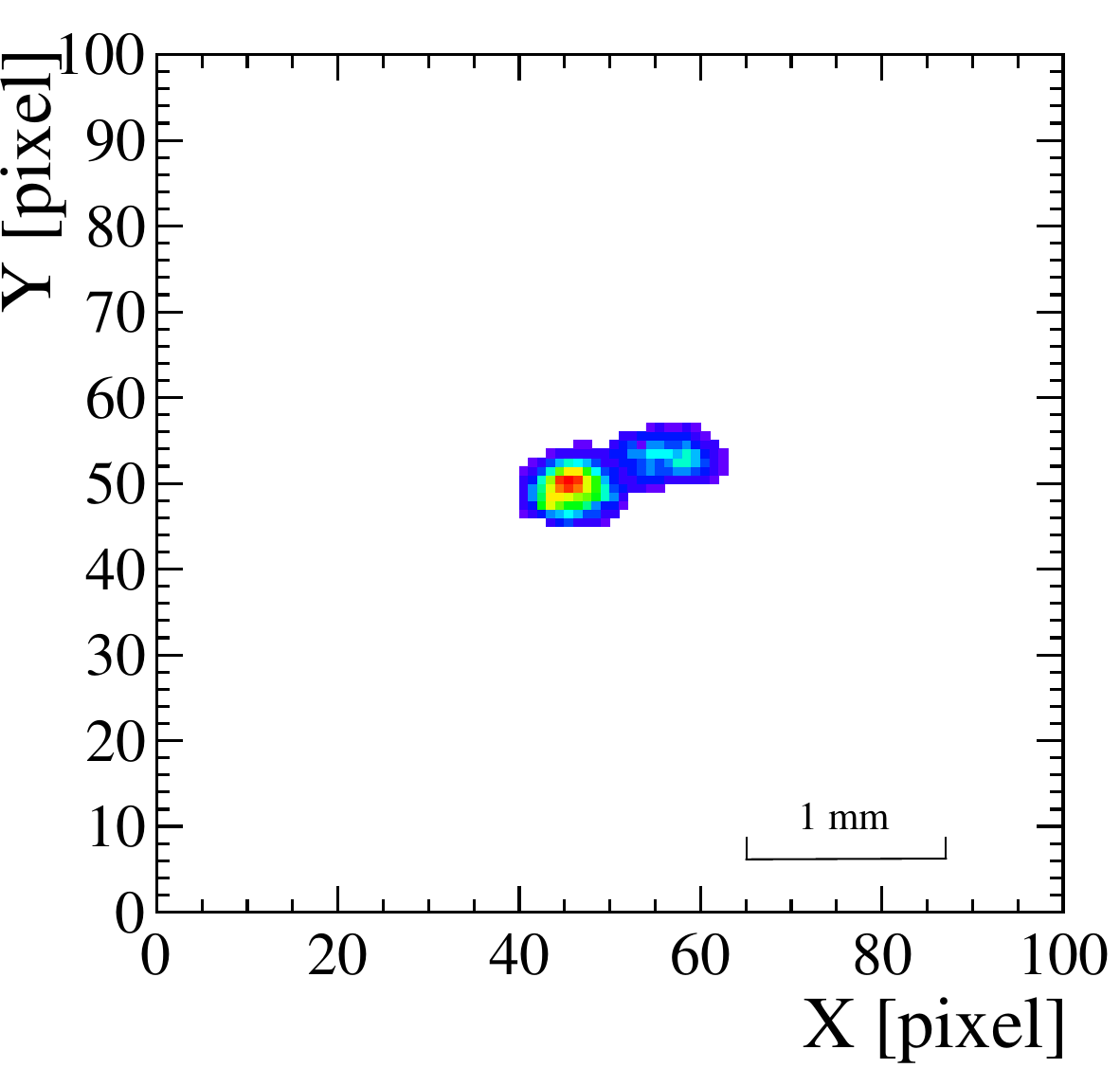}
    }
    \hfill
    \subfigure[]{
        \includegraphics[width=0.23\textwidth]{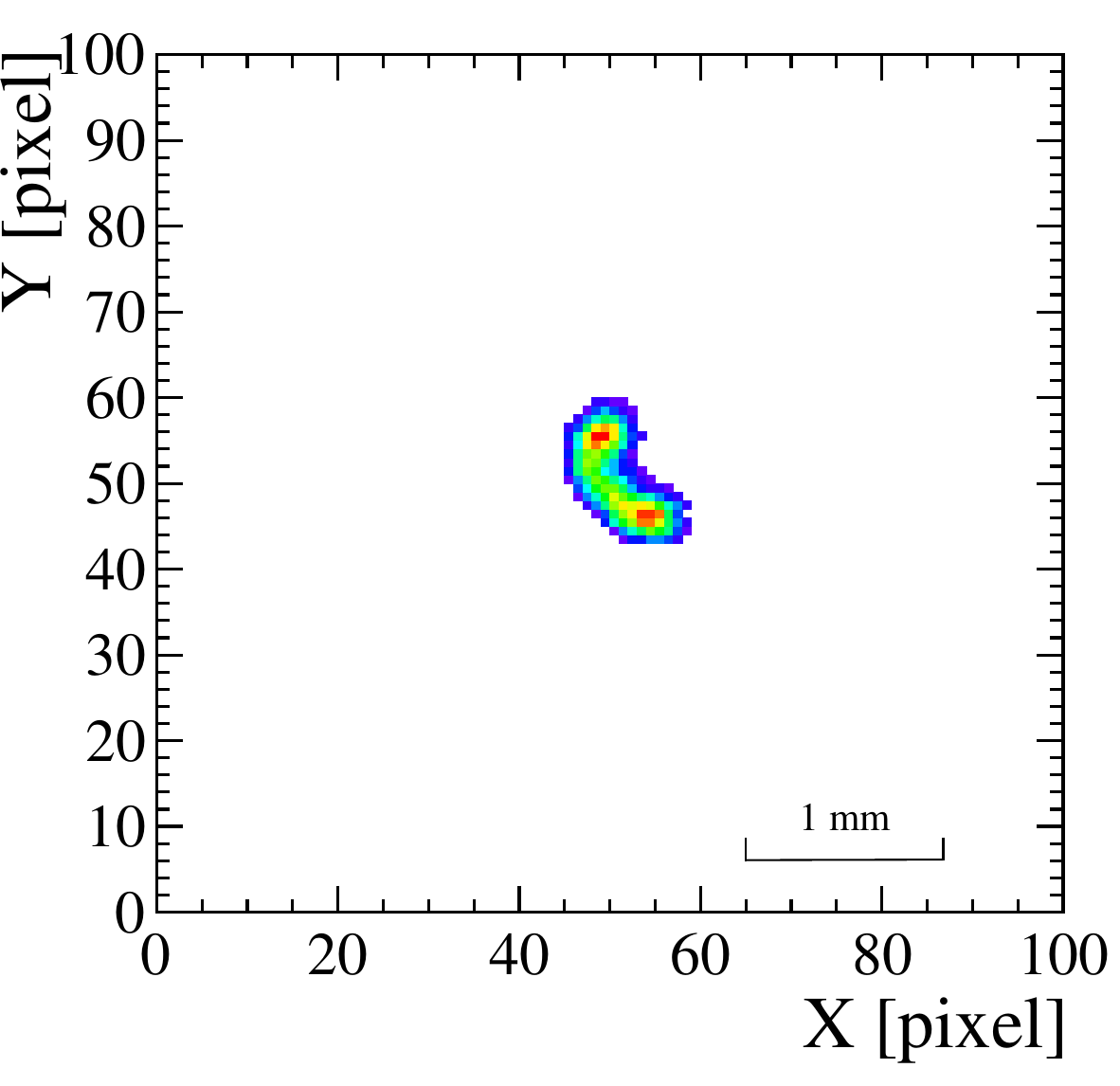}
    }
    
    \caption{Photoelectron tracks from a 5.90 keV X-ray source, measured in a 40\% He + 60\% DME gas mixture at 1 atm.}
    \label{fig:12}
\end{figure}

Fig.~\ref{fig:12} shows two typical track images obtained after background filtering. The pixel hue is proportional to the deposited charge, allowing clear observation of the complete photoelectron track and the Bragg peak at its end. The typical track length is approximately 1 mm. The key to extracting X-ray polarization information lies in accurately reconstructing the photoelectron emission direction.\par

\begin{figure}[htbp]
    \centering
    \includegraphics[width=0.48\textwidth]{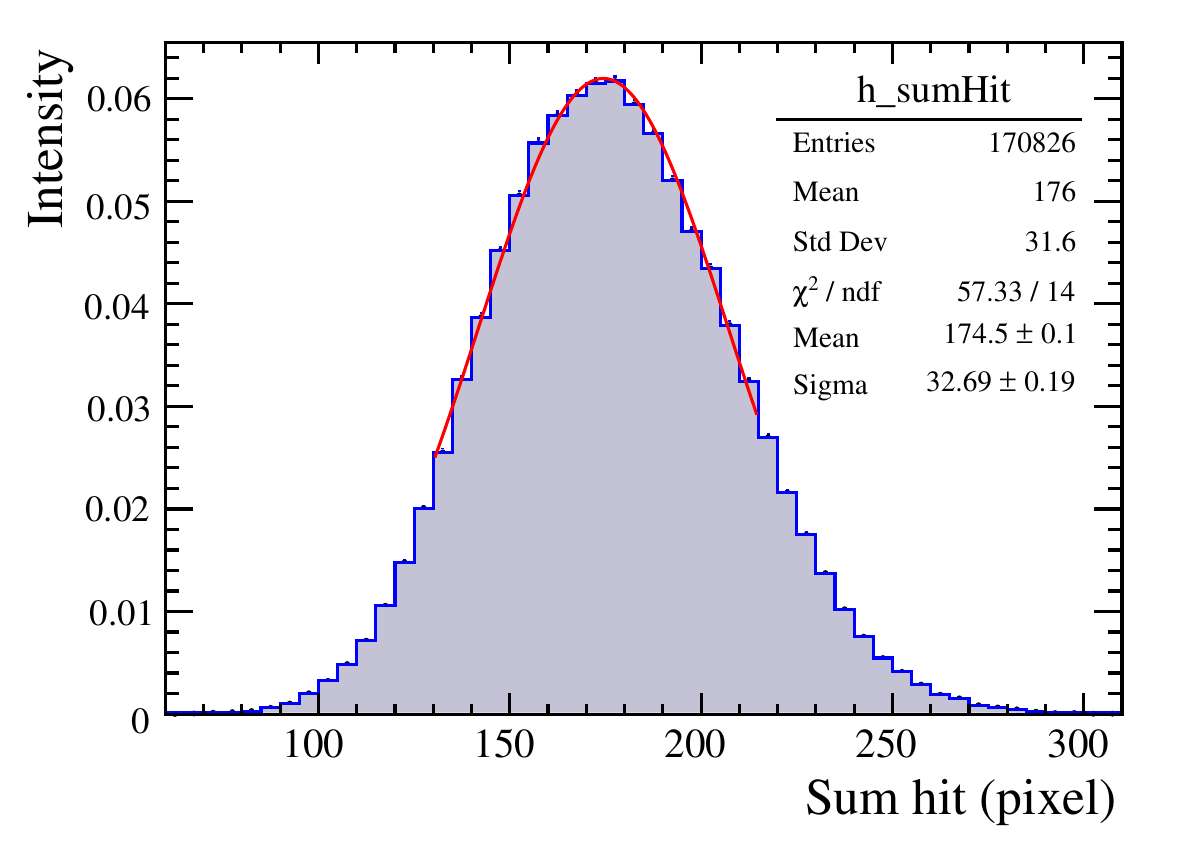}
    \caption{Pixel multiplicity distribution from photoelectron tracks induced by 5.90 keV X-rays, measured with Topmetal-L in a 40\% He + 60\% DME gas mixture at 0.8 atm.}
    \label{fig:13}
\end{figure}

To optimize the readout efficiency of Topmetal-L, we first employed a full-array rolling shutter scanning mode, accumulating over 170k X-ray events at 5.90 keV. By statistically analyzing the number of pixels triggered by each track on the chip and applying Gaussian fitting, we determined that a single event triggers approximately 176 pixels on average, as shown in Fig.~\ref{fig:13}. This statistical result provides a critical basis for setting the scanning interval in the subsequent sentinel readout mode. In the sentinel scanning mechanism, to ensure effective capture of track signals, the step size of the scanning interval must be smaller than the typical spatial extent of a track on the pixel array (N$_{pixel}$). Specifically, the interval should satisfy the relation: step size $< \sqrt{N_{pixel}}$. Based on the fitting result N$_{pixel}$ $\approx$ 176, the calculated square root is approximately 13.3. A scanning step size of 5 pixels, below the theoretical upper limit, was chosen to optimally balance detection efficiency and readout speed, thereby ensuring reliable detection of most valid track events.\par

\begin{figure}[htbp]
    \centering
    \includegraphics[width=0.48\textwidth]{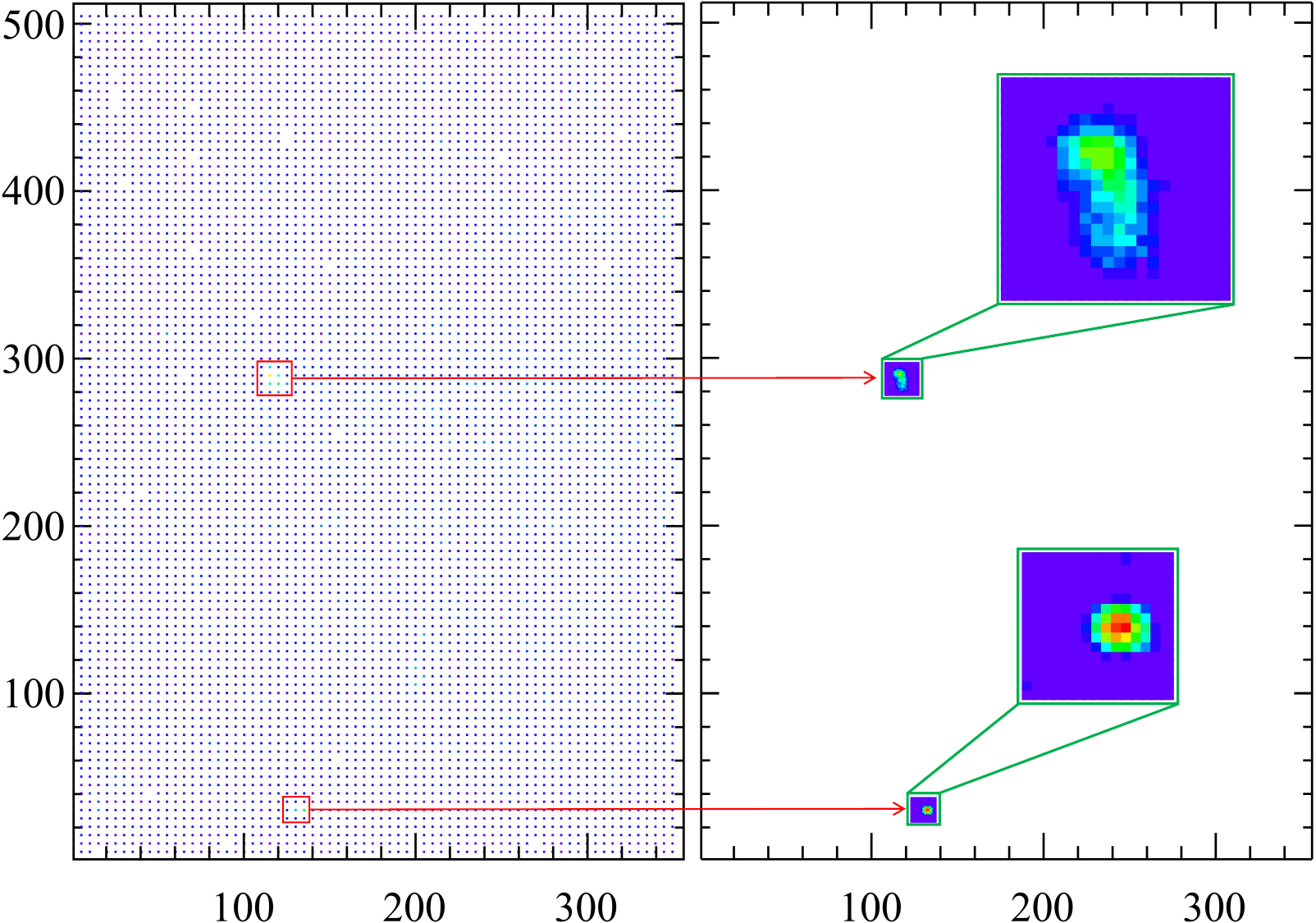}
    \caption{Readout performance of the sentinel scanning mode with a step size of 5 pixels for 5.90 keV X-rays in a 40\% He + 60\% DME gas mixture at 1 atm. The left panel shows one frame of the sentinel scanning, while the right panel shows one frame of the region scanning. A triggered pixel is marked by the red box, which led to the readout of a 15 $\times$ 15 pixel region and readout of a 20 $\times$ 20 pixel region during the region scanning.}
    \label{fig:14}
\end{figure}

When testing the sentinel readout mode, the trigger threshold is set to 200 ADC. When any pixel in the sentinel scanning frame exceeds threshold, the system automatically switches to the region scanning mode in the subsequent frame. The region scanning covers a square window centered on the address of the triggered pixel. Each sentinel is preset to monitor a 5 $\times$ 5 pixel area around it. To ensure track integrity, when a sentinel is triggered, the areas covered by its adjacent sentinels are also read out simultaneously. Fig.~\ref{fig:14} demonstrates the operational effectiveness of the sentinel readout mode through a case of dual-event triggering, showing two distinct activated regions: an upper region with a four-sentinel cluster and a lower region with a single sentinel. This results in the simultaneous readout of two regions with sizes of 20 $\times$ 20 and 15 $\times$ 15 pixels, respectively. With the scanning step size set to 5 pixels, the readout time for a single sentinel scanning frame is significantly reduced from 18.3 ms (required for full-array scanning) to 730 $\mu$s. Since region scanning typically involves only a few hundred pixels, the additional time overhead has a negligible impact on the overall system readout frame rate. This achieves a remarkable improvement in readout efficiency while maintaining track integrity. The raw data from the sentinel scanning frames, used solely for trigger determination, is discarded immediately after the trigger decision is made. This strategy significantly reduces the volume of invalid data at the source, providing substantial data compression benefits for large-scale pixel arrays. In space applications where bandwidth resources are extremely valuable, this approach effectively alleviates critical bottlenecks in the data transmission link, ensuring robust and stable long-term system operation.

\begin{figure}[htbp]
    \centering
    \includegraphics[width=0.45\textwidth]{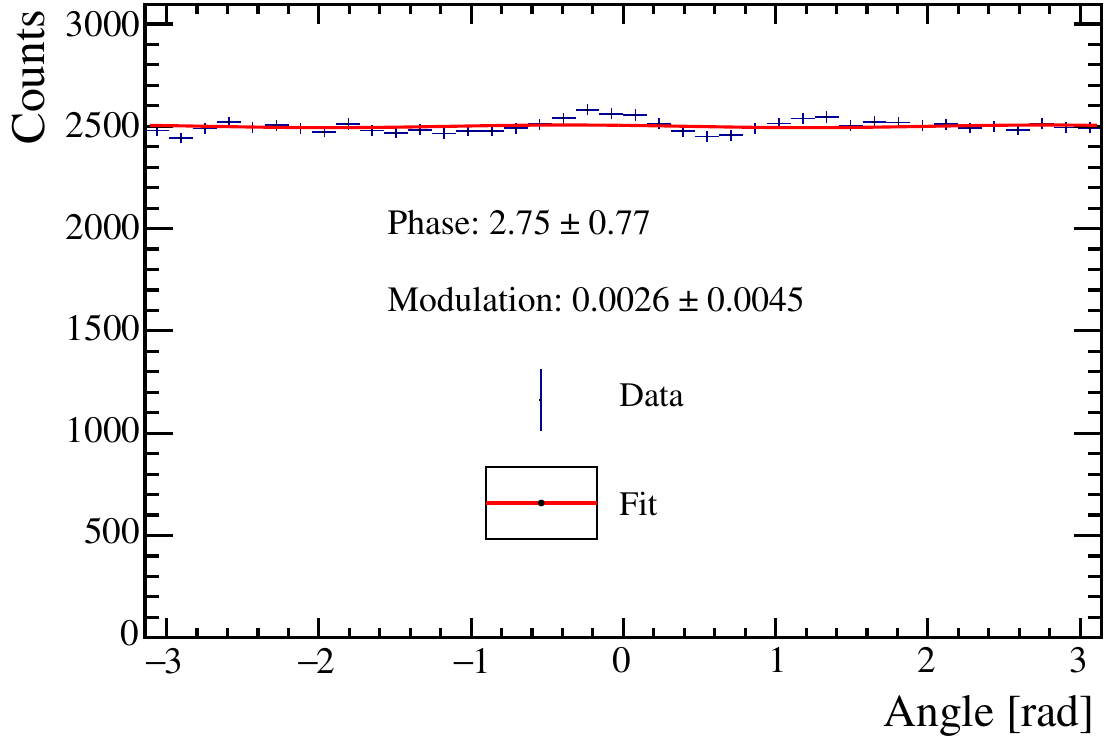}
    \caption{Modulation curve of 5.90 keV unpolarized X-rays measured with Topmetal-L as the charge sensor in a 40\% He + 60\% DME gas mixture at 0.8 atm.}
    \label{fig:15}
\end{figure}

\begin{figure}[htbp]
    \centering
    \includegraphics[width=0.48\textwidth]{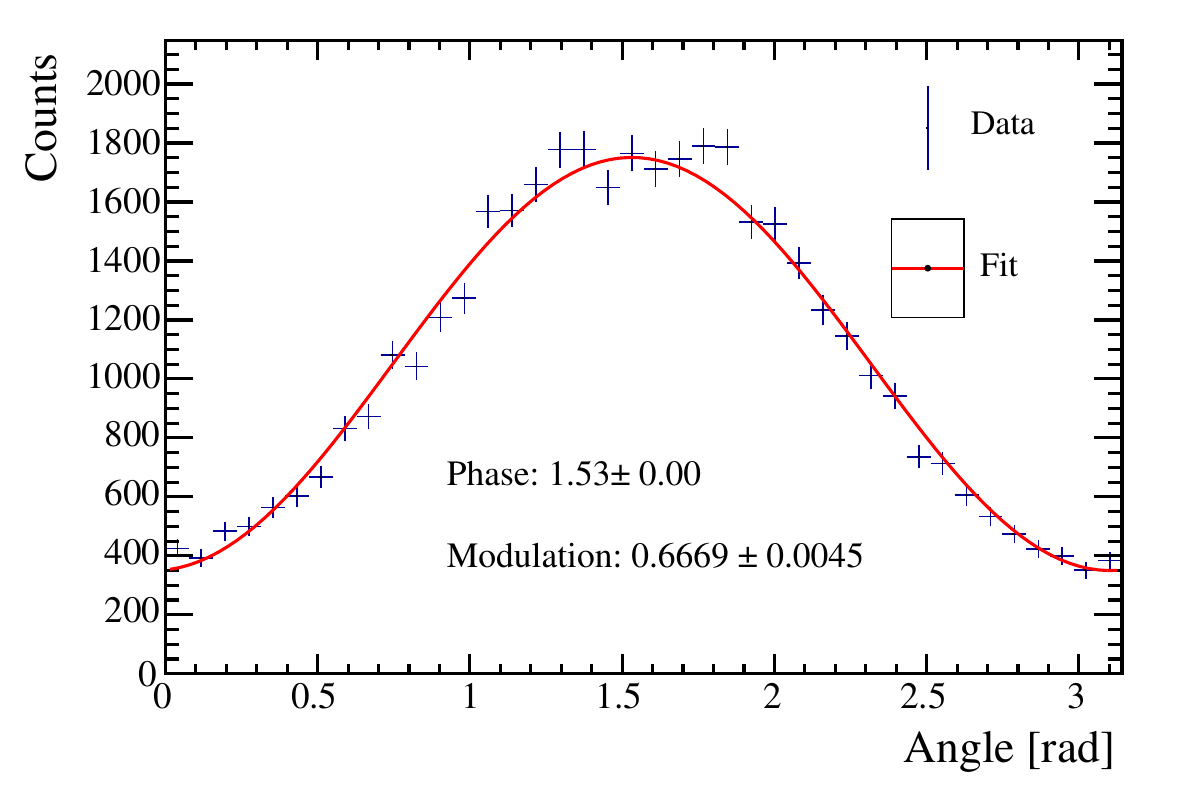}
    \caption{Modulation curve of 8.05 keV linearly polarized X-rays measured with Topmetal-L as the charge sensor in a 40\% He + 60\% DME gas mixture at 0.8 atm.}
    \label{fig:16}
\end{figure}

The modulation curve measured for 5.90 keV unpolarized X-rays by the detection system is presented in Fig.~\ref{fig:15}. The measured residual modulation factor is 0.26\% $\pm$ 0.45\%. Compared with the previous generation GMPD (using Topmetal-II$^{-}$, which exhibited a residual modulation factor of 1.96\% $\pm$ 0.58\%)\cite{29feng2023charging}, the present system reduces the residual modulation by approximately an order of magnitude, demonstrating significant progress in polarization measurement performance. This demonstrates the low systematic bias of Topmetal-L in polarization measurements. Fig.~\ref{fig:16} shows the modulation curve measured with the Topmetal-L detector for 8.05 keV linearly polarized X-rays, which were generated by a dedicated polarization platform based on the principles of Bragg diffraction and Fresnel reflection\cite{39xie2023variably}. The measured modulation factor of 66.67\% $\pm$ 0.45\% represents a remarkably high value, conclusively demonstrating the outstanding polarization measurement capability of Topmetal-L.\par

\begin{figure}[htbp]
    \centering
    \includegraphics[width=0.48\textwidth]{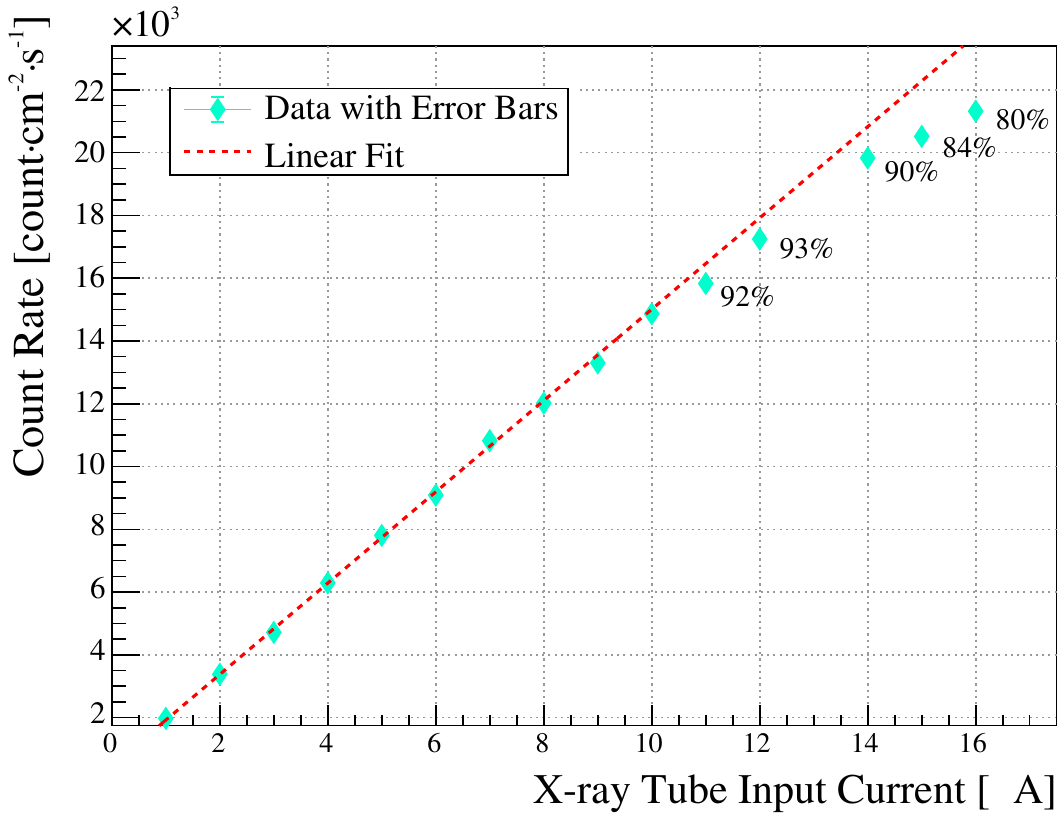}
    \caption{Count rate of Topmetal-L as a function of X-ray tube input current under an atmosphere of 40\% He + 60\% DME gas mixture at 1 atm (5.40 keV).}
    \label{fig:17}
\end{figure}

To evaluate the dynamic counting capability of Topmetal-L, tests utilized an X-ray tube with a precisely regulatable photon flux within a dedicated polarization platform. The output photon flux of the X-ray tube is proportional to the input current, allowing quantitative control of the incident flux by systematically adjusting the current, thereby enabling the measurement of the chip’s counting response under different flux levels. Fig.~\ref{fig:17} shows the measured count rate as a function of the X-ray tube input current for 5.40 keV X-rays. It can be observed that for input currents up to 10 $\mu$A, the count rate increases linearly with current, indicating accurate detector response without significant counting loss in this range. However, as the current increases further, the count rate gradually deviates from the linear trend, exhibiting clear nonlinear response and indicating the onset of detector saturation. Based on this response curve, it can be concluded that the Topmetal-L chip maintains a linear response up to a count rate of 15 k counts$\cdot\text{cm}^{-2}\cdot\text{s}^{-1}$. Beyond this threshold, data loss gradually intensifies. Nevertheless, this saturation count rate far exceeds the actual counting capability requirements of the LPD, demonstrating that Topmetal-L possesses sufficient counting margin for its intended application scenarios.\par

\begin{table*}[htbp]
\centering
\caption{Performance comparison}
\label{tab:2}
\renewcommand{\arraystretch}{1.5} 
\setlength{\tabcolsep}{11pt} 
\begin{tabular}{l|cccc} 
\toprule
 & \textbf{Topmetal-II$^{-}$} & \textbf{Topmetal-L} & \textbf{XPOL-I} & \textbf{XPOL-III} \\
\midrule
\textbf{Chip size [mm$^2$]} & 8 × 9 & 17 × 24 & - & - \\[0.5ex]
\textbf{Active area [mm$^2$]} & 6 × 6 & 16.02 × 23.04 & 15 × 15.2 & 15 × 15.2 \\[0.5ex]
\textbf{Pixel array} & 72 × 72 & 356 × 512 & 300 × 352 & 300 × 352 \\[0.5ex]
\textbf{Pixel density [pixels/mm$^2$]} & 145 & 494 & 462 (hexagon) & 462 (hexagon) \\[0.5ex]
\textbf{Power density [W/cm$^2$]} & 2.778 & 0.195 & 0.350 & - \\[0.5ex]
\textbf{Pixel gain [$\mu$V/e$^{-}$]} & - & 76.04 & 64.1 & 32.0 \\[0.5ex]
\textbf{ENC [e$^{-}$]} & 13.9 & 22.8 & 22.5 & 30 \\[0.5ex]
\textbf{Readout mode} & Rolling Shutter & Rolling Shutter / Sentinel & Self-trigger & Self-trigger \\[0.5ex]
\textbf{Time cost} & 2.6 ms/frame & 0.73 ms/frame @ Sentinel & 1 ms/event & 0.15 ms/event @ 2.5 keV \\
\bottomrule
\end{tabular}
\end{table*}

Based on experimental results from the previous-generation LPD prototype, Topmetal-L incorporates targeted upgrades optimized for space applications. As summarized in Table~\ref{tab:2}, the new-generation chip shows significant improvements across multiple key performance parameters compared to Topmetal-II$^{-}$ and achieves performance levels competitive with the XPOL-I and XPOL-III\cite{40minuti2023xpol}, while offering distinct advantages in power consumption and sensitive area. The two architectures are driven by distinct scientific objectives: the XPOL series, utilizing self-triggered region-of-interest readout, is architected for pointed observations of focused sources, whereas Topmetal-L—with its large sensitive area, low power density, and sentinel scanning scheme—is specifically designed for wide-field surveys of transient phenomena such as gamma-ray bursts. Notably, Topmetal-L's sentinel readout allows multiple distinct particle tracks to be triggered and processed within a single frame, as illustrated in Fig.~\ref{fig:14}. Despite this different approach, its full-chip event throughput remains similar to that of XPOL-III.

\section{Summary and outlook}
\label{sec:summary}

This work has successfully designed, implemented, and validated the Topmetal-L, a large-area, low-noise charge-sensitive pixel sensor custom-developed for the POLAR-2/LPD space-borne polarimeter. Fabricated in a 130 nm CMOS process, the chip features a 356 $\times$ 512 pixel array with a pitch of 45 $\mu$m, where each pixel incorporates a 26 $\times$ 26 $\mu$m$^2$ charge collecting window. The Topmetal-L sensor simultaneously provides energy and position information of deposited charges, achieving an active area of 3.69 cm$^2$ and an equivalent noise charge of 22.8 e$^-$ at a total power consumption of 720 mW. To meet the high-count-rate requirements for transient source observations, an innovative sentinel scanning readout scheme was proposed, effectively resolving the conflict between the large-scale pixel array and limited downlink bandwidth, thus reducing the effective frame readout time to 730 $\mu$s. \par
System-level tests demonstrate that the detection prototype based on Topmetal-L exhibits high polarimetric purity (residual modulation factor: 0.26\% ± 0.45\% at 5.90 keV) and sensitivity (modulation factor of 66.67\% ± 0.45\% for 8.05 keV polarized X-rays), with a maximum counting rate capability reaching 15 k counts$\cdot\text{cm}^{-2}\cdot\text{s}^{-1}$. The upper limit of the linear count response significantly exceeds the requirements of the LPD mission. All core performance metrics confirm that Topmetal-L is a viable technological solution that meets the next-generation space astronomy demands for large sensitive area, low power consumption, and high event throughput.\par
Subsequent efforts will focus on engineering implementation and system integration, including comprehensive environmental adaptability tests such as radiation hardening, mechanical vibration, and thermal vacuum cycling to evaluate the chip's long-term reliability under extreme space conditions. Furthermore, extending this sensor architecture to other particle imaging applications based on gas detectors will be an important direction for future exploration.\par

\bibliographystyle{ieeetr}
\small\bibliography{sample}

\end{document}